\documentclass[pra,aps,twocolumn,showpacs]{revtex4}

\usepackage{epsfig,pstricks}
\usepackage{graphicx}
\usepackage{dcolumn}
\usepackage{bm}
\usepackage{bbm}
\usepackage{amsmath}
\usepackage{amssymb}
\usepackage{amsfonts}

\def\id{{\rm 1\kern-.22em l}}

\begin{document}
\title{Hidden order in bosonic gases confined in one dimensional optical lattices}
\author{L. Amico$^{(a)}$, G. Mazzarella$^{(b)}$, S. Pasini$^{(c)}$, and  F.S. Cataliotti$^{(d)}$}

 \affiliation{(a)\,
MATIS-INFM $\&$ Dipartimento di Metodologie Fisiche e Chimiche
(DMFCI), Universit\'a di Catania, viale A. Doria 6, 95125 Catania,
ITALY}
\affiliation{(b)
Dipartimento di Fisica "G.Galilei", Universit\'a di Padova, via F.
Marzolo 8, 35131 Padova, ITALY} \affiliation{(c) Lehrstuhl f\"{u}r
Theoretische Physik I, Technische Universit\"{a}t Dortmund,
 Otto-Hahn Stra\ss{}e 4, 44221 Dortmund, GERMANY}
\affiliation{(d) Dipartimento di Energetica  $\&$ LENS,
Universit\'a di Firenze, via N. Carrara 1, 50019 Firenze, ITALY }

\begin{abstract} We analyze the effective Hamiltonian arising from a
suitable power series expansion of the overlap integrals of Wannier
functions for confined bosonic atoms in a $1d$ optical lattice. For
certain constraints between the coupling constants, we construct an
explicit relation between such an effective bosonic Hamiltonian and
the integrable spin-$S$ anisotropic Heisenberg model. Therefore the
former results  to be  integrable by construction. The field theory
is governed by an anisotropic non linear $\sigma$-model with singlet
and triplet massive excitations; such a result holds also in the
generic non-integrable cases. The criticality of the bosonic system
is investigated. The schematic phase diagram is drawn. Our study is
shedding light on the hidden symmetry of the Haldane type for one
dimensional bosons.  \end{abstract}


\maketitle

\section{Introduction}
Systems of cold atoms trapped in optical lattices provide a
remarkable tool to simulate quantum many body physics in engineered
quantum systems \cite{review.cold,BlochRMP}.
In this context new perspectives are
provided by the possibility to handle atoms or molecules with large
dipole moment \cite{Stuhler}. For these systems, the Bose-Hubbard
models can be employed\cite{Baranov-Zaccanti},
provided that processes arising from the long range interaction are
 considered. The corresponding 'extended' Bose-Hubbard
Hamiltonian can display a variety of quantum phase transitions
between superfluid states to Mott Insulator (MI) ones (f.i. see
\cite{Saro}).

Recent investigations by Altman and coworkers indicate that a
strongly interacting bosonic system can display a further gapped
phase
 with 'exotic' order \cite{Altman}. Based on numerical
analysis of a Bose-Hubbard type model, such phase was demonstrated
to be characterized by a non local {\it hidden} order of the Haldane
type. Originally discovered in spin systems \cite{Haldane,Hidden},
Haldane phases are believed to play a crucial role in many different
contexts including QCD \cite{Azcoiti} and High--T$_C$
superconductivity \cite{Super}. Recently the hidden order was
noticed also in different types of electronic
insulators\cite{Rosch,Nussinov}.

The Haldane order in the bosonic systems was evidenced by
rephrasing the physical concepts  provided
to study of the hidden order for spin systems. Although Altman
and coworkers constructed a variational scenario in order to
interpret their results, it would be desirable to trace how
the bosonic hidden order emerges from the Haldane order in  spin
systems. This is a major purpose of the present work.

 Based on exact
methods we construct a precise relation between the bosonic and spin
hidden orders. Relying on it the quantum criticality of the bosonic
system is investigated. We construct the non-linear-$\sigma$ model
(NL$\sigma$M) describing the system at long wave lengths. The phase
diagram is investigated within the saddle point approximation of the
NL$\sigma$M. Our results are shedding  light on the structure of the
hidden order in bosonic systems, suggesting that the bosonic
excitations in the Haldane Insulator (HI) are of  triplet nature.

The paper is organized as follows.
In Sec. II we derive the model
Hamiltonian from the  microscopic dynamics of
a dilute system of interacting dipolar bosons. The
emerging Hamiltonian defines a type of extended Bose-Hubbard model of interacting
bosons in the lowest Bloch band.
In section III, we establish the  mapping between the bosonic model and  the
integrable higher spin-$S$ $XXZ$ Hamiltonian\cite{ZAMO-FADDEEV} (see Eqs.(\ref{FTTdeformedone})-(\ref{smallanistropybis})).
Because of the integrability of the spin model, the
bosonic Hamiltonian is integrable by construction as long  the microscopic parameters
entering into bosonic Hamiltonian fulfill certain relations (see Eq.(\ref{constraints})).
Relying on this result, the quantum criticality of the bosonic system
is investigated in the Sec. IV where we construct the non-linear-$\sigma$ model
(NL$\sigma$M) describing the system at long wave lengths.
The phase
diagram is constructed, beyond integrability,  within the saddle point approximation of the
NL$\sigma$M (see Fig.(\ref{phasediag}); we  note that the results  are  not restricted by Eq.(\ref{constraints})).
In section IV we  discuss and suggest experimental protocols to detect the
Haldane insulating phase. Finally  we  draw our conclusions.

\section{The bosonic model Hamiltonian}
We start by analyzing the order of magnitude of the amplitudes
involved in the $1d$ lattice Hamiltonian for trapped bosonic atoms;
we include dipole-dipole interaction. We will obtain an effective
model with density-density interaction and higher order hopping
processes.

The general Hamiltonian for the atomic (of mass $m$) gas reads
$H=H_0+H_{int.}$ with
\begin{eqnarray}
H_0=-\int d \vec{r}\hat{\Psi}^{+}(\vec{r})\left
[\frac{\hbar^2}{2m}\nabla^2-V(\vec{r})\right ]
\hat{\Psi}(\vec{r}), \label{Hamiltonian} \\
H_{int}=\int d\vec{r}d\vec{r}^{'}
\hat{\Psi}^{+}(\vec{r})\hat{\Psi}^{+}(\vec{r}^{'})V_{int.}(\vec{r}-\vec{r}^{'})
\hat{\Psi}(\vec{r}^{'})\hat{\Psi}(\vec{r})\; ,\nonumber
\end{eqnarray}
where $V$ results from a combination of harmonic confinement with
the optical lattice: $V=V_{harm} + V_{latt}$. We consider a
'cigar-shaped' configuration $V_{harm}=m\omega^2 (x^2+\gamma^2
y^2+\gamma^2 z^2)/2$ where the harmonic potential has a frequency
$\omega$ along the $x$ direction and is much more confined of an
anisotropy factor $\gamma\gg1$ in the $y,z$ directions.
The $1d$ lattice, $V_{latt}= sE_r \sin^2 (\pi x/a)$, is arranged
along $x$.
$E_r=(\hbar \pi)^{2}/2a^2m$ is the photon recoil energy, $a$ is the
lattice spacing and $s$ measures the optical lattice depth in terms
of $E_r$.

The interaction potential $V_{int.}=V_{sr}+V_{dd}$ contains two
terms: an 'onsite' short range potential
$V_{sr}(\vec{r})=4\pi\hbar^2 a_{BB}\delta(\vec r)/m$ characterized
by the s-wave scattering length $a_{BB}$; and a 'long range'
anisotropic dipole-dipole potential
$V_{dd}(\vec{r}-\vec{r}^{'})={\mu_{0}\mu^{2}} ({1-3\cos^2
\theta})/({{4\pi} |\vec{r}-\vec{r}^{'}|^3})$,  $\mu$ being the
atomic magnetic dipole ($\mu_0$ is the vacuum magnetic
permeability), and $\theta$ being the angle of $\vec{r}-\vec{r}^{'}$
with the dipoles orientation.

The bosonic field operator can be realized through  Wannier
functions $w(\vec{r}-\vec{r}_i)$ localized around $\vec{r}_{i}$: $
{\hat{\Psi}(\vec{r})=\sum_{i} b_{i}w(\vec{r}-\vec{r}_i)}$, where
$b_{i}$  annihilates   a boson at the lattice site $i$. In this
formalism the Hamiltonian reads ${H =  \sum_{i,j} t_{ij} b^\dagger_i
b_j+\sum_{ijkl}t_{ij;kl} b_i^{\dagger} b_j^{\dagger} b_k b_l}, $
where $t_{ij}$ and $t_{ij;kl}$ are the standard Wannier functions
integrals\cite{review.cold}. They  can be expanded as power series
of the so called 'lattice attenuation parameter' $\varepsilon\doteq
\exp(-a^{2}/4l_{opt}^{2})=\exp[-(\pi/2)^2 \sqrt{s}]$, with
$l_{opt}=a/(\pi s^{1/4})\ll a$ \cite{BlochRMP};
$w(\vec{r}-\vec{r}_i)$ are assumed to factorize:
$w(\vec{r}-\vec{r}_i)=w(x-x_i)w(y)w(z)$ and approximated as
${w(v)={(\pi^{-1/4}{l_{v}^{-1/2}})} {\exp(-{v^2}/ 2l_v^{2})}}$,
$v=x,y,z$, with $l_x\doteq l_{opt}$ and
$l_y=l_z=l_\bot=(\hbar/m\gamma \omega)^{1/2}$.

The effective lattice (grand canonical) Hamiltonian emerging from
the analysis above is $H=E_r H_b$ with
\begin{eqnarray}
\label{particlehamiltonian} &&{H}_b=-\kappa\sum_i n_i -t\sum_{i}
{b}^{\dagger}_{i}{b}_{i+1} + U_0\sum_i
{n}_i({n}_{i}-1) \nonumber \\
&&+U_1 \sum_{i} {n}_i{n}_{i+1}
-t_c\sum_{i} ({n}_{i}+{n}_{i+1}) {b}^{\dagger}_{i}{b}_{i+1} +  \nonumber \\
&&t_{p} \sum_{i} ({b}^{\dagger}_i)^2 ({b}_{i+1})^2 +\dots \,
,\end{eqnarray} where ${n}_{i}={b}^{\dagger}_{i}{b}_{i}$ is the
number operator. Terms in the ellipses involve higher powers in
$\varepsilon$ (see also \cite{Santos}).
The first two terms, with $\kappa$ being the chemical potential and
$t$ the nearest neighbor hopping (in units of $E_r$), are the only
low order terms from $H_0$ since the matrix elements $t_{ij}$
decrease as $\varepsilon^{|i-j|^{2}}$.
$U_{0}=(2/\pi)^{3/2}{a_{BB}a}s^{1/4}/l_{\bot}^{2} $ is the on-site
interaction neglecting the renormalization due to $V_{dd}$ (that is
of order $\varepsilon^2$). Defining $I_{dd}=m\mu_0 \mu^2/(2
a\pi^3\hbar^2) $ the contributions of $V_{dd}$ to the integral
$t_{ii+1;ii+1}$, the coupling constants in
(\ref{particlehamiltonian}) read
\begin{eqnarray}
\label{couplingconstants} t_c &=&U_{0}
\varepsilon^{3/2}+I_{dd}\varepsilon \;
,\nonumber\\
U_{1}&=& 2t_p=\left (U_{0} +I_{dd}\right ) \varepsilon^{2}\; ,
\end{eqnarray}
We comment that the (form of the) second quantized Hamiltonian
(\ref{particlehamiltonian}) is not affected by the gaussian
approximation of the Wannier functions which might only modify the
expression of the coupling constants in (\ref{couplingconstants}).


Besides the density-density interaction $U_1$, $H_b$ provides for
correlated $t_c$ and bosonic pair $t_p$ nearest-neighbors hopping
processes. It is important to note that experimentally there are
different ways of manipulating the relative strength of the coupling
constants e.g. it is possible to change the optical lattice
parameters \cite{Eckholt}, or use appropriate Feshbach resonances
\cite{fattori} to separately tune $I_{dd}$ and $U_0$,
\textit{independently} from  the expansion in terms of the parameter
$\varepsilon$.
This allows for the possibility of exploring a large portion of
phase space.


Below we demonstrate that  the $1d$ bosonic Hamiltonian
(Eq.(\ref{particlehamiltonian})) is obtained as large $S$ limit of
the integrable spin model generalizing the $XXZ$ model to higher
spin \cite{Matsuda-Liu}. 
This will prove that the model (\ref{particlehamiltonian}) is
integrable (see however the discussion below
(\ref{smallanistropybis})).

\section{The integrable Heisenberg model for higher spin}
Integrable
$XXZ$ models for higher spin \cite{ZAMO-FADDEEV,SOGO} were
intensively studied in the framework of the quantum inverse
scattering method \cite{KOREPIN-BOOK}. The  Hamiltonian ($E_r$ is assumed as  the energy's untit) is obtained
as logarithmic derivative of the transfer matrix \cite{BYTSKO}
\begin{equation}
\label{FTTdeformedone}
 H_{XXZ}=-\sum_{i}
\psi_\alpha[J(\alpha)_{i,i+1}+1]+\psi_\alpha(1)\;,
\end{equation}
where
\begin{eqnarray}
\label{psialpha}
&&\psi_\alpha [x]=\Gamma'_\alpha(x)/\Gamma_\alpha(x) \nonumber\\
&&\Gamma_{\alpha}(x)~=~(1-e^\alpha)^{1-x} \prod_{n=0}^{\infty}
{[1-e^{\alpha(n+1)}]}/{[1-e^{\alpha (n+x)}]}\; .\nonumber\\
\end{eqnarray}
reducing to the
ordinary gamma function for $\alpha\rightarrow 0$. The quantity
$J(\alpha)$ is related to the (coproduct $\Delta$ of) Casimir
operator
$
C_S=S^+ S^- +{\sinh\alpha S^z\sinh \alpha (S^z+1)}/{\sinh^2\alpha}
$ of the quantum algebra
$su_\alpha(2)$ underlying the integrability of the theory:
\begin{eqnarray}
\label{deltacsone}
\Delta C_S=\frac{\sinh \alpha J(\alpha)
\sinh \alpha (J(\alpha)+1)}{2\sinh^2\alpha} \;. \end{eqnarray}
We
remark that the anisotropy $\alpha$ enters into
(\ref{FTTdeformedone}) deforming both the gamma function and the
'representation' $J(\alpha)$ of $su_\alpha(2)$.

The limit of large $S$, with small $\alpha S$ is interesting for us.
The first 'effect' of such a  limit
is that the $\psi_\alpha(x)$ in
(\ref{FTTdeformedone}) reduces to the ordinary digamma function
$\psi(x)=\Gamma^{'}(x)/\Gamma(x)$. In such a limit the quantity $J(\alpha)$ can be obtained explicitely by resorting
to the explicit expression of $\Delta C_S$
\begin{eqnarray}
\label{deltacstwo}
\Delta C_S^{(i,i+1)} &=& e^{\alpha S_{z}}\bigg[\frac{1}{2} S^{+}_{i}S^{-}_{i+1}
+\frac{1}{2} S^{-}_{i}S^{+}_{i+1}+\nonumber\\
&&\makebox{\hspace*{-1.4cm}}+\sinh \alpha S^{z}_{i} \sinh \alpha S^{z}_{i+1} \frac{\cosh \alpha S \cosh
\alpha (S+1)}{\sinh^2 \alpha}\nonumber\\
&&\hspace*{-1.4cm}+\cosh \alpha S^{z}_{i} \cosh \alpha S^{z}_{i+1}\frac{\sinh \alpha S \sinh
\alpha (S+1)}{\sinh^2 \alpha}\bigg]e^{-\alpha S_{z}}\; .\nonumber\\
\end{eqnarray}
By retaining the terms up to the second order in $\alpha S$  we obtain
\begin{eqnarray}
\Delta C_S^{(i,i+1)} &=& S^x_iS^x_{i+1}+S^y_iS^y_{i+1}+\lambda
S^z_iS^z_{i+1} \nonumber\\
&+& \frac{\alpha^2}{2}\big((S^z_i)^2+(S^z_{i+1})^2\big)+S(S+1)\;,
\end{eqnarray}
where $\lambda=1+\alpha^2 k $ with $k=S(S+1)+1/2$. By exploiting its
relation with $\Delta C_S$, $J(\alpha)$ can be written as  $1/S$ perturbative expansion: $J(\alpha)_{i,i+1}=A(\alpha)_{i,i+1}+B(\alpha)_{i,i+1}S $ to finally give
\begin{eqnarray}
J(\alpha)_{i,i+1} =&&
 2 S \nonumber \\
&&+1/2[-3\alpha^2/8+ (1-3\alpha^2/8) \left
( \hat{b}^{\dagger}_{i} \hat{b}_{i+1}+h.c.\right ) \nonumber \\
&& -(1+9\alpha^2/8)
\left (\hat{n}_{i} + \hat{n}_{i+1}\right )]
\end{eqnarray}
 where we used the
Holstein-Primakoff realization of $su(2)$ \cite{Auerbach}.
The bosonic model (\ref{particlehamiltonian}) is obtained first
employing  the Stirling formula for the asymptotics of the digamma
function $\psi(x)\approx \ln (x)$, then by expanding the equation
resulting from (\ref{FTTdeformedone}) at second order in $1/S$:
\begin{eqnarray}
\label{smallanistropybis} H_{XXZ} =H_b +const.\;,
\end{eqnarray}
where $const.=N\ln  [S (2+3\alpha^2/4)]-N\psi(1)$.
Translational invariance has been assumed. The spin $S$, and the
anisotropy $\alpha$ are obtained by comparison of
(\ref{smallanistropybis}) with (\ref{particlehamiltonian}):
$S={t}/({2U_0+U_1})$, $\alpha^2={2(2U_0-U_1)}/({3U_1}) $,  with the
constraints
\begin{eqnarray}
\label{constraints}
&& t_c=\frac{ U_1}{2}\;, \quad  t_p={\frac{ 2t_c-U_0}{2}} \nonumber \\
&& \frac{(2U_0+U_1)^2}{8t^2}=\frac{8(2 U_0-3  U_1)^2}{(8t-14  U_0+11 U_1)^2}  \nonumber \\
&& 8t^2 U_1= (2U_0+U_1)^2 
\end{eqnarray}
We note that only one
parameter results to be adjustable in 
(\ref{particlehamiltonian}) achieved from
(\ref{FTTdeformedone}); we remark that the restrictions (\ref{constraints}) can be achieved by  tuning the relative strenght of contact versus dipole interactions. The resulting one-parameter-$H_b$ is
integrable by construction (see also \cite{BOGOLIUBOV}); the exact
solution will be studied elsewhere. In the isotropic case
$\alpha=0$, $H_b$ is obtained from the
Faddeev-Takhadjan-Tarasov-Babujian model \cite{FTT}. In this
case: $U_1= U_0=t^2=2t_c=4t_p$ \cite{AmicoKorepin}.

The ground state of (\ref{FTTdeformedone}) is a singlet:
$S^z_{tot}=0$. For imaginary $\alpha$ the spectrum is gapless. For
real $\alpha$ it was proved that the excitations are gapped
\cite{SOGO}. Given the relation (\ref{smallanistropybis}), it is
intriguing to study the low energy spectrum of $H_{XXZ}$ in the
limit of large $S$. This is what we are going to do by exploiting
the NL$\sigma$M.

\section{Non linear sigma model}

As customerly, the large $S$ limit
is combined with  the continuous limit $a \rightarrow
0$.
The action ${\cal S}=i S \omega[\Omega] +\int d\tau dx {\cal H}[\Omega]
(\tau,x) $ is obtained  within the spin coherent states
path integral formalism: ${\cal H}[\Omega]=\langle \Omega| H| \Omega \rangle$ and $\omega[\Omega]$ is the Berry phase\cite{Auerbach}.

The semiclassical continuous Hamiltonian arising from gradient expansion of
(\ref{FTTdeformedone}) is ${\cal H}= \langle \Omega(x,\tau) |\Delta
C_S|\Omega (x,\tau)\rangle+\dots $ where the ellipses indicate terms
with higher powers of $a$ and  $1/S$ (or combinations of thereof).
${\cal H}$ has $\lambda-D$ form with $D=\alpha^2$; the gradient expansion
procedure spoils the integrability of the lattice model
(\ref{smallanistropybis}), since the $t_c$ and $t_p$ terms turn out
of higher order in $a$\cite{Affleck-continuos}; nevertheless
integrability manifests in the field theory in form of the
restriction $\lambda=1+kD$, with $k=S(S-1/2)$, arising from the lattice theory.
Generic
values of $\lambda$, $D$ can be considered through  the $1/S$
expansion of the $\lambda-D$ lattice model (that is not integrable),
providing a different parametrization of the spin parameters in 
bosonic terms:
\begin{equation}
S=\frac{4t+1}{8}\,, \; \lambda=U_1 \,,\; D=U_0\; .
\end{equation}
The large scale behavior of the system is captured
by the large $S$ realization of the coherent state spin variables
in terms of staggered magnetization
$\vec{n}_{j}(\tau)$ and quasi-homogeneous
$\vec{l}_{j}(\tau)$ fluctuating fields through the Haldane mapping \cite{Haldane}
\begin{equation}
\label{Haldane_mapping}
  \langle \Omega|\vec{S}_{j}(\tau)|\Omega \rangle =(-1)^{j}\vec{n}_{j}(\tau)
\sqrt{1-\frac{|\vec{l}_{j}(\tau)|^2}{S^2}}+\frac{\vec{l}_{j}(\tau)}{S},
\end{equation}
where  $\vec{n}_{j}(\tau)^2=1$ and
$\vec{l}_{j}(\tau) \cdot \vec{n}_{j}(\tau)=0$.
The NL$\sigma$M description of (\ref{FTTdeformedone}) is obtained  following a variation of the procedure originally
adopted  in  \cite{Haldane,Campos,PasiniPh.D} to deal with the
$S=1$, $\lambda-D$ model. Due to the anisotropy of the $\lambda-D$ model, we separate the staggered magnetization and its fluctuation in their perpendicular and transversal components: $\vec{n}_{j}=(\vec{n}_\bot,n_z)$ and $\vec{l}_{j}=(\vec{l}_\bot,l_z)$.
In order to obtain the Lagrangian of the NL$\sigma$M one needs to integrate out the $\vec{l}$ field in the Hamiltonian together with the terms coming from the Berry phase. After taking the continuum limit one finds
\begin{eqnarray}
\label{nlsm}
{\cal L}={\frac{a^2}{2}} \left[S^2|\partial_x {\bf n}_\perp|^2 +
{\frac{c_\perp}{2}} |\partial_\tau {\bf n}_\perp |^2 \right ] +&&  \\
&& \hspace*{-4.5cm} {\frac{a^2}{2}}
\left[S^2\lambda \left (\partial_x { n_z}\right )^2+
{\frac{c_z}{2}} \left (\partial_\tau n_z \right )^2 \right ] +M S^2 n_z^2 \;,  \nonumber
\end{eqnarray}
with $M=1+S(S-1/2) U_0-U_1$, plus the topological phase
associated with the N\'{e}el field $\vec{n}=({\bf n}_\perp,n_z)$
\begin{equation}\label{T_n}
 T=\frac{1}{4\pi}\int
d\tau dx\
\vec{n}\cdot\left(\partial_{\tau}\vec{n}\times\partial_x\vec{n}\right).\end{equation}
$T$ can assume only integer values for it is the winding number of the mapping $\vec{n}:\mathbbm{R}_{comp}\mapsto\mathbbm{S}^2$ which is classified by the second homothopy group $\pi_2(\mathbbm{S}^2)=\mathbbm{Z}$. $T$ contributes in the partition function as $e^{2i T\pi S}$, with $T\in \mathbbm{Z}$, so that it does not contribute for integer spins.
In Eq.(\ref{nlsm})
the N\'{e}el field satisfies the constraint $|\vec{n}|^2=1$.
The coefficients are
\begin{eqnarray}
\label{cz_cperp}
c_\bot &=& 2\frac{1-n_z^2(1-U_1)}{(\mu-Mn_z^2)(2-M n_z^2)} \nonumber \\
c_z &=& \frac{\mu-2n_z^2(1-U_1)}{(\mu-Mn_z^2)(2-M n_z^2)},
\end{eqnarray} with $\mu=1+U_1+kU_0$. The coefficients
provide an additional
interaction between the components of the N\'{e}el field.
For $M=0$, Eq. (\ref{nlsm}) is the \textit{sum} of an
$O(2)$ NL$\sigma$M in the field $\mathbf{n}_\bot$ and a scalar model in $n_z$.
The former describes a free Gaussian model with a bosonic compactified field which is
known to be a conformal field theory with central charge $c=1$;
the latter is also integrable.
\begin{figure}
\begin{center}
 \includegraphics[width=0.48\textwidth]{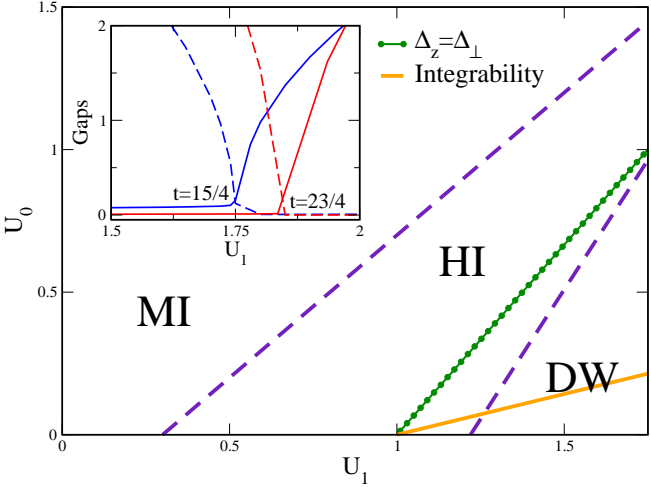}
\end{center}
\caption{Saddle point phase diagram for the NL$\sigma$M. Based on the behavior of the gaps
(inset), the different phases are identified through the indications
we reported  in the text; the small offsets in the decay of
$\Delta_z$ and  $\Delta_\bot$ is artifact of the saddle point
approximation. The plot corresponds to $t=15/4$. Along the orange
continuos line $U_1=1+k U_0$ the lattice model
(\ref{particlehamiltonian}) is integrable. The green
dotted-continuos line indicates the degeneration of the gaps:
$\Delta_z=\Delta_\bot$ (a similar phenomenon was evidenced for
$S=1$\cite{Campos,PasiniPh.D,Botet}). {\it Inset}: Behavior of the
gaps $\Delta_z$ (dashed lines) and $\Delta_{\bot}$ for $U_0=1$ and
for different $t$ (leftmost blue lines correspond to $t=15/4$ while
red lines to $t=23/4$). For increasing $t$ the HI is shrunk.
\label{phasediag}}
\end{figure}
One of the most striking features of the Haldane phase is that the gap is provided by a triplet
structure\cite{Haldane,Zamo-Zamo}. The masses of the particles
$\Delta_\bot$ and $\Delta_z$ belong
 to the sectors $S_z=0$ and $S_z=\pm 1$ respectively.
For the case $S=1$
$\Delta_z$ and  $\Delta_\bot$ play the role of the excitations
of a conformal field theory  with  $c=1/2$ and
 $c=1$ respectively\cite{Campos}.
Eventhough the  Haldane order in the bosonic systems was evidenced through a non
local 'order parameter' defined in analogy with the spin string
order parameter\cite{REVIEW-SPIN}, the phase diagram
can be studied also by analysing the  low energy properties of the
system and specifically the particle-hole $\delta{\cal E}_c$ and the
neutral $\delta{\cal E}_n$ energy gaps\cite{Altman}.
We shall see that $\Delta_z$ and $\Delta_\bot$ play the role of
 $\delta{\cal E}_c$ and $\delta{\cal E}_n$ respectively.

To obtain such quantities from the continous field theory  we treat (\ref{nlsm}) by saddle point approximation ($S=1$).
The method assumes that $<n_z^2>:=\zeta$. This means that $\|(\delta
n_j)^2-\delta n_j \delta n_{j+1}\|\ll 1 $ with $\delta n=n-\bar{n}$,
indicating that only sparse charge-fluctuations around 'diluted DW
states' (so called 'zero defects states'\cite{Gomez-Santos}) are
considered. The constraint on the field $\vec n$ is taken into account by introducing a uniform Lagrangian multiplier $\eta$
\begin{eqnarray}
\label{nlsm_eta}
{\cal L}&=&{\frac{a^2}{2}} \left[\frac{v_\bot}{g_\bot}|\partial_x {\bf n}_\perp|^2 +
 \frac{|\partial_\tau {\bf n}_\perp |^2}{v_\bot g_\bot} \right ] +M S^2 n_z^2   \nonumber\\
&+& {\frac{a^2}{2}}
\left[\frac{v_z}{g_z}\lambda \left (\partial_x { n_z}\right )^2+
\frac{\left (\partial_\tau n_z \right )^2}{v_z g_z} \right ] + \eta(x,\tau)
(|{\mathbf{n}}|^2-1),\nonumber\\ \end{eqnarray}
where $v_\bot/g_\bot=S^2$, $v_z/g_z=S^2\lambda$ and $1/(v_i g_i)=c_i/2$ with $i=\{\bot, z\}$.
There are two  poles in the propagators of the effective
action, providing the expressions in terms of $\zeta$ for the gaps in the $z$ and in the perpendicular direction, respectively $\Delta_z$ and
$\Delta_\bot$, to be determined
selfconsistently\cite{PasiniPh.D}. In order to derive the expression for the gaps we need the calculate the propagator. We substitute in the action the Fourier transform of the field $\vec n$ and of the Lagrange multiplier $\eta$, respectively $\vec{\tilde n}$ and $\tilde \eta$.  After performing a Gaussian integration on the field
$\tilde{\mathbf n}$, the effective action becomes:
\begin{equation}
 \label{no_n_action}S[\tilde \eta]= \sum_{q,n,j}\left\{
\ln (K^{-1})^{jj}\right\}(q,n;q',n') -\tilde\eta(0,0)\ ,\end{equation} where $(K^{-1})^{jj}$ are the diagonal entries of the inverse propagator
\begin{eqnarray}
 \nonumber\label{propag}&&(K^{-1})^{jj}= \\ \nonumber
&&\frac{1}{\beta L}
\left\{\frac{\delta_{nn'}\delta_{qq'}}{2g_j v_j}(\Omega_n^2+q^2
v_j)-\frac{i\tilde\eta}{\beta L}(\Delta q,\Delta n)+\mu\ \delta_{j3}\right\},\\ \end{eqnarray}
with $\Delta q:=q-q'$, $\Delta n:=n-n'$ and $j=\{1,2,3\}$. The notation $v_j$, $g_j$ stands for $v_\bot$ $g_\bot$ if $j=\{1,2\}$ or $v_z$ and  $g_z$ if $j=3$. $\Omega_n$ are the Matsubara-Bose frequencies, $\Omega_n=2\pi n/\beta$.
$(K^{-1})^{33}$ provides an expression for $\zeta$.

With the ansatz $\tilde\eta(q,n)=(\beta
L)\delta_{n0}\delta_{q0}\ \eta $, we can derive the saddle point
equation ($\partial S/\partial \eta
 =0$). For low temperatures
($\beta \to \infty$) and in the thermodynamic limit ($L\to\infty$)
one finds
\begin{equation}
 \label{selfconsist_eqn} \left\{\begin{array}{l} 1 \
=\frac{g_\bot}{\pi}\ln \left[ {\Lambda}{\xi_\bot}
+\sqrt{{\Lambda^2}{\xi_\bot^{2}}+1}\ \right]+\zeta \\ \zeta
=\frac{g_z}{2\pi}\ln\Bigl[
\frac{\Lambda}{\sqrt{\xi_z^{-2}+\nu_z^{-2}}}
+\sqrt{\frac{\Lambda^2}{\xi_z^{-2}+\nu_z^{-2}}+1}\ \Bigr]
\end{array}\right.\end{equation}
where the ultraviolet cutoff $\Lambda$ is a free parameter to be
determined  by fitting the gap to the known value for the isotropic point $\Delta_{isotr}=0.41048$ where $\xi_\bot=\xi_z$.
The notation stands for $\xi_\gamma^{-2}:=\frac{2 g_\gamma
\eta}{v_\gamma}$ with $\gamma=\{\bot ,z\}$, $\xi_z^{-2}:=\xi_\bot^{-2}/\lambda$ and
$\nu_z^{-2}:=\frac{2 M g_z }{v_z}$. The expressions for the gaps are
\begin{equation}
 \label{gaps}\left\{\begin{array}{l}
\Delta_\bot = v_\bot\xi_\bot^{-1}\\
\Delta_z = v_z\sqrt{\xi_z^{-2}+\nu_z^{-2}}\ .
\end{array}\right.\end{equation}
We can solve numerically the self-consistency equations
(\ref{selfconsist_eqn}) for different values of $U_1$,
$U_0$ and of the spin $S$ (or equivalently $t$).
From the numerical study of the gaps one can draw a (qualitative) phase diagram (see Fig.\ref{phasediag}).

We see that $\Delta_z$ and $\Delta_\bot$ display a behavior that
is very similar to $\delta{\cal E}_c$ and $\delta{\cal E}_n$
respectively (see Fig.\ref{phasediag}). In such a view, the
phenomenology of the bosonic excitations in the HI would arise from
the triplet nature of $\Delta_z$ and $\Delta_\bot$. In particular,
the line $\delta{\cal E}_c=\delta{\cal E}_n$  that was evidenced
numerically in Ref.\cite{Altman} can be interpreted as the
degeneracy of the triplet excitations of the field theory
(\ref{nlsm}) (that includes the $O(3)$ model
$\lambda=0$~\cite{Zamo-Zamo}).

Further insights on the criticality of the bosonic system
can be obtained by adapting the results for the spin $S$, $\lambda-D$
model\cite{Schulz}. Specifically,
the onset to the HI-density wave phase is suggested to be of the Ising type (second order),
$c=1/2$  and it is characterized by $\Delta_z=0$ with $\Delta_\bot\neq 0$;
the MI-HI phase transition is with $c=1$ and it is caused by $\Delta_z\neq 0$
with $\Delta_\bot= 0$.
Similar
behavior of $\delta{\cal E}_c$ and $\delta{\cal E}_n$
 was noticed in
\cite{Altman} (except that also $\delta{\cal E}_c=0$ was displayed to vanish at the  MI-HI transition).
The density wave (DW) phase-MI transition is predicted
to be of the first order; the $c=1/2$ and $c=1$ lines meet in a point that is governed by the continuos limit of the $\alpha=0$ integrable theory, with
 $c=3S/(1+S)$, $S=1/(2t)$\cite{Affleck-cft}.
The phase diagram of the system is summarized in Fig.\ref{phasediag}.

\section{Experimental feasibility}
Although a detailed analysis would go far beyond the scope of the
present paper, we would like to sketch a possible method  that could
serve for detecting the HI experimentally  (complementing the
observations in \cite{Altman}). The basic idea relies on the
detection of the atomic current tracing back to the early
experimental evidence of MI phase \cite{Greiner}: the MI
conductivity is probed by applying a washboard potential to the
lattice. A resonance in the atomic conductivity appears when the
tilt between adjacent sites reaches the energy gap $U_0$ {\it i.e.}
when it is resonant with the particle-hole pairs excitation energy.
Because of the peculiar solitonic non local order\cite{Gomez-Santos}
of the HI the resonance peak in the conductivity is narrowed, as
already noticed with a different experimental situation (parametric
resonance) by \cite{Altman}.
Placing the bosonic chain in a washboard potential, the  distinct
feature of the HI phase would be a dependence of the atomic current
on the length of the chain (this is the analog  of the diffusive
spin transport evidenced in Haldane
compounds\cite{Haldane-diffusive}). We observe that, being an incompressible phase, the HI is
expected to be robust to the parabolic confinement whenever the
induced total energy offset (between center and the trap edges) is
of the order of the energy gap. Following the numerical indications
provided in \cite{Altman} such a gap is estimated to be of the same
order of magnitude of the MI one: $\Delta\simeq t$. Therefore we can
estimate the magnitude of the allowed harmonic confinement to be $m
\omega^2 x_M^2<t=2s\varepsilon E_r$, $x_M$ being the size of the
condensate along the optical lattice. It would be interesting to
trace the analogous for the HI of the "wedding cake" structure that
appears in MI in presence of strong confinement.

Finally, ring shaped (with circumference $L$) optical lattices
with twisted boundary
conditions\cite{AOC} could be
exploited to evidence the HI in an expansion experiment
(we note that the energy offset  due to harmonic confinement
can be minimized in this case).
In fact, adapting the results obtained for the spin diffusion in
twisted Heisenberg  integer spin rings\cite{Kopietz},
the atomic density current would display a characteristic parametric
dependence on the boundary twist: a sawtooth like behaviour
for correlation lenght $\xi\gg L$, and  it would be exponentially suppressed,
 with sinusoidal oscillations    for $\xi\ll L$.
This could be a fingerprint of the Haldane gap for (finite)
ultracold atomic systems.

\section{Conclusions}
By exploiting  a suitable expansion of the matrix elements in terms
of the lattice attenuation parameter $\varepsilon \ll 1$ we derived
an effective  model for bosonic atoms in a $1d$ lattice
(\ref{particlehamiltonian}). Additional terms enter the Hamiltonian
respect to the standard Bose-Hubbard model. For certain choice of
the coupling constants, the model results to be integrable through a
mapping with the spin-$S$-$XXZ$ integrable model (see
Eq.(\ref{smallanistropybis})). We note that the possibility of
independently tuning the onsite-interaction relative to the
density-density interaction, is available in the bosonic model
Hamiltonian only in presence of dipole-dipole interactions. This is
precisely what allows degenerate quantum gases to explore different
portions of the phase diagram beyond the MI region.


The direct relation between
the spin and bosonic pictures are exploited to investigate the
critical properties of the bosonic systems. From the present work,
it is suggested that the correct context is provided by the spin
$S$, $\lambda$-$D$ model. The HI is investigated by studying the
continuos field theory arising from gradient expansion of the lattice model: the Lagrangian has the form of a NL$\sigma$M with
further interactions between the components of the Neel field (see (\ref{nlsm}), (\ref{cz_cperp})).
We comment
that the terms $t_p$ and $t_c$ entering the
(\ref{particlehamiltonian}) do not contribute to the NL$\sigma$M
formulation indicating that such terms are irrelevant for the criticality we
studied. Such an effect is a physical manifestation of the Affleck
analysis \cite{Affleck-continuos}. The difference between integrable
and non integrable lattice theories is reflected in a different
parametrization of the coupling constants entering the NL$\sigma$M. Interestingly enough the integrable parameterization
of the continous model (\ref{nlsm}) we found should be tightly related to the so called pricipal chiral fields\cite{principal}
indicating an emergent $SU(2)\times SU(2)$ symmetry of the bosonic Haldane phase.
We notice however that the phase diagram of the system (\ref{nlsm}) is investigated, {\it beyond the integrability},
by means of the saddle point approximation; the integrable case might need separate discussion. The Haldane phase is
found for any finite range of the interaction. This should be
relevant for experiments where the scattering length can be tuned
around zero thus evidencing the interaction between light--induced
dipoles \cite{fattori}.

{\acknowledgments}
We thank M.A. Martin-Delgado, A. Pronko, A. Sedrakyian,
G. Sierra, and F. Sols for very useful discussions. L.A.
acknowledges support from MEC (FIS2007-65723); FSC acknowledges support
from the EU through project CHIMONO; G.M. acknowledges support 
from 'Fondazione CARIPARO'.

\end{document}